\documentclass[a4paper,11pt]{article}
\usepackage{pos}

\title{Light (anti)nuclei production via coalescence in electron–proton collisions: from HERA to EIC}
\ShortTitle{Light (anti)nuclei production via coalescence}

\author*[a,b]{Niccolò Ciavarelli}
\author[b]{Francesco Noferini}
\author[a,b]{Francesca Bellini}

\affiliation[a]{Department of Physics and Astronomy A.Righi, University of Bologna,\\
  Viale Carlo Berti Pichat, 6/2, Bologna, 40126, Italy}

\affiliation[b]{INFN Sezione di Bologna, Italy}

\emailAdd{niccolo.ciavarelli@studio.unibo.it}

\abstract{Light (anti)nuclei production provides a sensitive probe of the particle-emitting source and of the coalescence mechanism in high-energy collisions. In this work, a coalescence model implemented as an afterburner to \textsc{Pythia8} is validated against pp data from ALICE and ep data from ZEUS. The validated framework is then applied to ep collisions at EIC energies to predict deuteron and antideuteron production within the ePIC acceptance. The results indicate sizeable yields and a clear matter--antimatter asymmetry in the proton-going direction, making this measurement a promising observable for studying coalescence and proton structure at the EIC.}

\FullConference{The 33rd International Workshop on Deep Inelastic Scattering and Related Subjects (DIS2026)\\
4 - 8 May 2026\\
Bologna, Italy\\}

\begin{document}
\maketitle

\section{Introduction}

Light nuclei and antinuclei are produced in high-energy collisions through the coalescence of nucleons sufficiently close in phase space at kinetic freeze-out~\cite{Kapusta1980,Scheibl1999,Blum2017}. Their production depends on the nucleon-emitting source and the nuclear wave function, providing a sensitive probe of the hadronization environment. A proton and a neutron can coalesce into a deuteron, while the corresponding antinucleons can form an antideuteron. If a bound state is not formed, the relative momentum distribution of the outgoing nucleons retains information on the particle-emitting source and can be studied through femtoscopic correlations~\cite{Ciavarelli2026}.

The coalescence probability depends on the nucleon phase-space distribution and, in particular, on the pair average transverse momentum $k_T$, the source size, and the spatial extent of the nucleus~\cite{Scheibl1999,Bellini2019}. Increasing the source size reduces the probability of forming a bound state because the overlap with the nuclear wave function becomes smaller. This behaviour is supported by ALICE measurements showing a clear dependence of light-nucleus production on the emitting-system size~\cite{ALICE2023}.

Deep-inelastic electron--proton and electron--nucleus collisions provide a complementary environment to study this mechanism. Their cleaner and better-constrained initial state makes them suitable for testing the universality of coalescence across collision systems. Deuteron and antideuteron production in the proton-going region can also probe the matter--antimatter asymmetry of the nucleon source and may provide sensitivity to proton and nuclear parton distribution functions.

In this work, a coalescence model is validated using pp and ep measurements and applied to predict deuteron and antideuteron production at the future Electron--Ion Collider.

\section{Simulation strategy and coalescence model}

To simulate light (anti)nuclei production, \textsc{Pythia8} is used to generate collision events, followed by a custom afterburner implementing the coalescence model~\cite{Mahlein2023}. The afterburner operates event-by-event on final-state nucleons and evaluates their probability to form a bound state.

The nucleon-emitting source is described by a Gaussian distribution multiplied by a plane wave,
$
\psi_{\mathrm{source}}(\vec{r},\vec{k}^{\,*})
=
A
\exp\left(-\frac{r^{2}}{8R_{0}^{2}}\right)
\exp\left(i\vec{k}^{\,*}\cdot\vec{r}\right),
$
where $A$ is the normalization amplitude, $\vec{r}$ the relative nucleon distance, and $R_{0}$ the source radius. The pair relative momentum in its rest frame is
$
k^{*} =
\frac{\left|\vec{p}_{1}^{\,*}-\vec{p}_{2}^{\,*}\right|}{2},
$
where $\vec{p}_{1}^{\,*}$ and $\vec{p}_{2}^{\,*}$ are the nucleon momenta in the pair rest frame.

The (anti)deuteron ground-state wave function is obtained by modelling the strong nucleon--nucleon interaction with a spherical square-well potential,
$
V(r)=-V_{0}\,\Theta(R_{D}-r),
$
where $R_{D}$ is the (anti)deuteron radius and $V_{0}$ the square-well depth. For each $R_{D}$, $V_{0}$ is fixed to reproduce the (anti)deuteron binding energy, $E_{\mathrm{BE}}\simeq 2.2$~MeV.

The coalescence probability is obtained from the projection of the source wave function onto the (anti)deuteron ground-state wave function,
$
P_{\mathrm{coal}}(k^{*})
=
\frac{3}{8}
\left|
\left\langle
\psi_{\mathrm{source}}(\vec{r},\vec{k}^{\,*})
\middle|
\psi_{\mathrm{D}}(\vec{r})
\right\rangle
\right|^{2}.
$

The factor $3/8$ accounts for the spin--isospin configuration required to form a spin-1 (anti)deuteron from two spin-$1/2$ nucleons. The free parameters are the source radius $R_{0}$ and the (anti)deuteron radius $R_{D}$, fixed by tuning to experimental data.

\section{Results}

\subsection{Validation}

The model was first tuned using pp collision data and validated against ALICE measurements~\cite{ALICE2020}. The observable is the coalescence parameter $B_{2}$, corresponding to the $A=2$ case of the general parameter $B_{A}$. Assuming identical proton and neutron momentum distributions,
$
    E_{A}\frac{d^{3}N_{A}}{d^{3}p_{A}}
    =
    B_{A}
    \left(
        E_{p,n}\frac{d^{3}N_{p,n}}{d^{3}p_{p,n}}
    \right)^{A}
    \bigg|_{p_{p}=p_{n}=p_{A}/A}.
$

For deuterons and antideuterons, $A=2$, and $B_{2}$ provides an experimental measure of bound-state formation, allowing comparison across collision systems.

The validation uses $10^{9}$ pp collisions at $\sqrt{s}=13.6$~TeV generated with \textsc{Pythia8} Monash 2013 and processed with the coalescence afterburner. The source radius $R_{0}$ and the (anti)deuteron radius $R_{D}$ are treated as free parameters and fixed through tuning.

\begin{figure}[t]
    \centering
    \begin{minipage}[t]{0.49\textwidth}
        \centering
        \includegraphics[width=\textwidth]{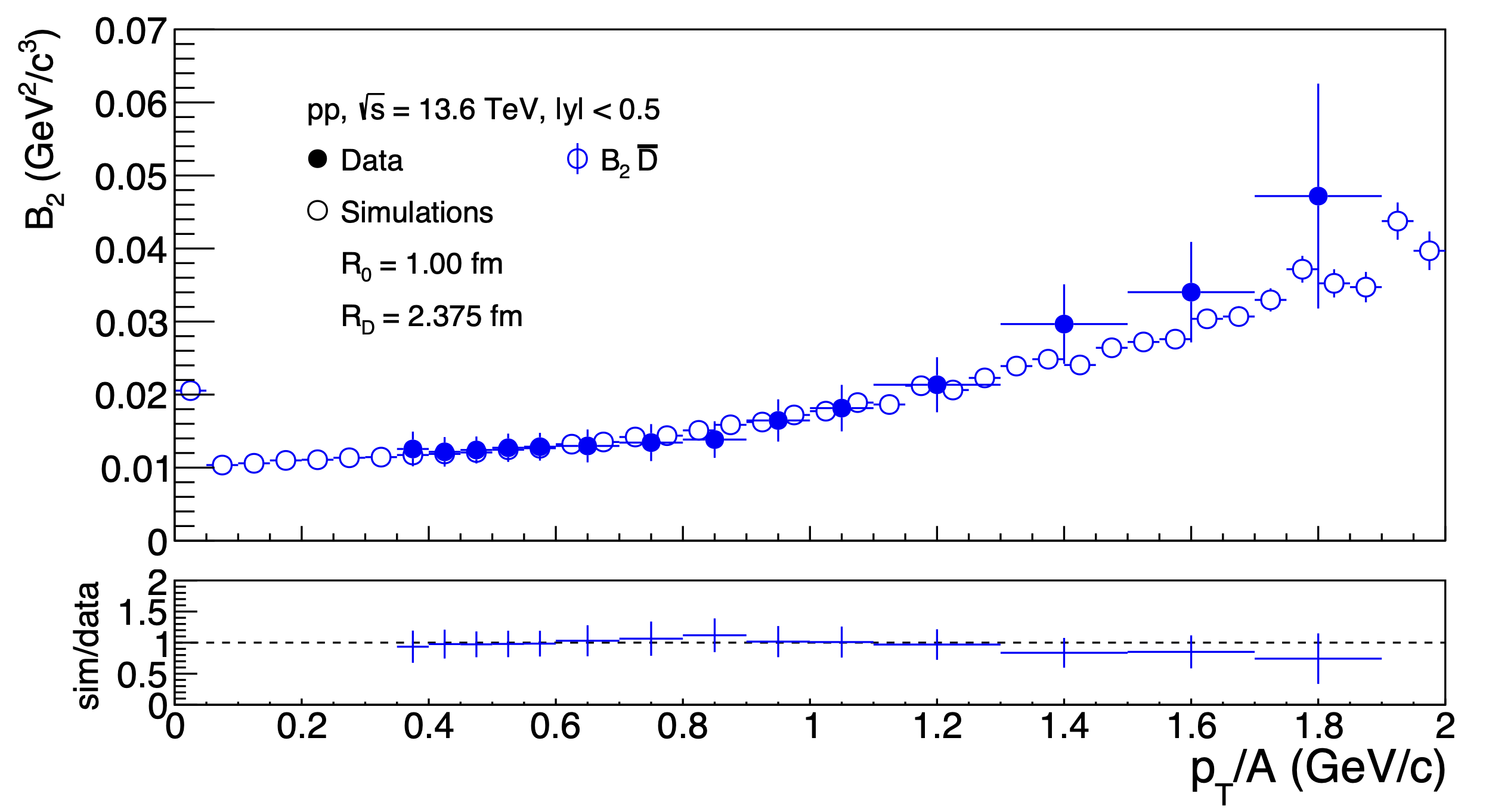}
        \caption{Coalescence parameter $B_{2}$ for antideuteron production in pp collisions at $\sqrt{s}=13.6$~TeV as a function of $p_{\mathrm{T}}/A$. Simulations are obtained from $10^{9}$ pp collisions generated with \textsc{Pythia8} Monash 2013 and processed with the coalescence afterburner, using $R_{0}=1.00$~fm and $R_{D}=2.375$~fm. Filled markers represent ALICE data~\cite{ALICE2020}, while open markers show the model results. The lower panel reports the simulation-to-data ratio.}
        \label{fig:b2validation}
    \end{minipage}
    \hfill
    \begin{minipage}[t]{0.49\textwidth}
        \centering
        \includegraphics[width=\textwidth]{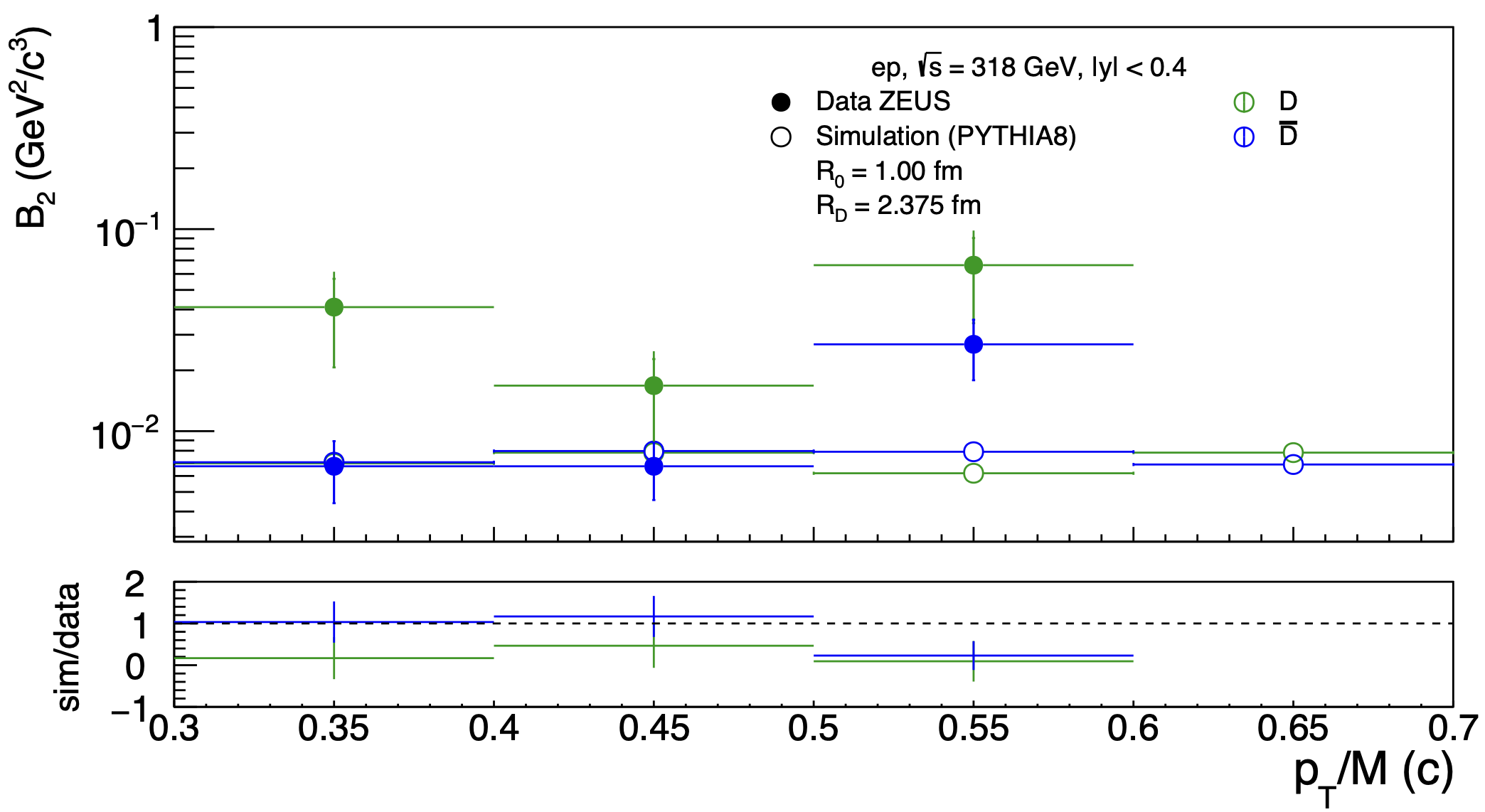}
        \caption{Coalescence parameter $B_{2}$ for deuteron and antideuteron production in ep collisions at $\sqrt{s}=318$~GeV as a function of $p_{\mathrm{T}}/M$. Filled markers represent ZEUS data~\cite{ZEUS2007}, while open markers show predictions obtained with \textsc{Pythia8} and the coalescence afterburner using $R_{0}=1.00$~fm and $R_{D}=2.375$~fm. The lower panel reports the simulation-to-data ratio.}
        \label{fig:b2hera}
    \end{minipage}
\end{figure}

As shown in Fig.~\ref{fig:b2validation}, $B_{2}$ increases with $p_{\mathrm{T}}/A$. The model reproduces the ALICE measurements within experimental uncertainties over the full measured range, with the simulation-to-data ratio compatible with unity.

The model was then tested against HERA ep data using ZEUS measurements~\cite{ZEUS2007}. The pp-tuned parameters, $R_{0}=1.00$~fm and $R_{D}=2.375$~fm, are used to evaluate $B_{2}$ for deuteron and antideuteron production at $\sqrt{s}=318$~GeV.

As shown in Fig.~\ref{fig:b2hera}, the predictions are consistent with ZEUS data within experimental uncertainties and reproduce the overall magnitude of the measured $B_{2}$ values for deuterons and antideuterons. This supports the applicability of the coalescence framework to ep collisions.

The comparison with HERA is particularly relevant for testing the universality of coalescence across collision systems. Previous ALICE measurements compared the coalescence parameter across several systems, including pp, pA, $\gamma$p, and ep collisions~\cite{ALICE2018}. HERA provides an important ep reference, although with limited statistical precision. The future Electron--Ion Collider (EIC), with $\mathcal{L} = 5.33~\mathrm{fb}^{-1}$ in the first year of data taking, will allow a significantly more precise measurement.

\subsection{Predictions}

Building on the validated framework, the model is applied to ep collisions at EIC energies to predict deuteron and antideuteron production as a new observable for studying light-nuclei formation in deep-inelastic scattering.

\begin{figure}[t]
    \centering
    \begin{minipage}[t]{0.59\textwidth}
        \centering
        \includegraphics[width=\textwidth]{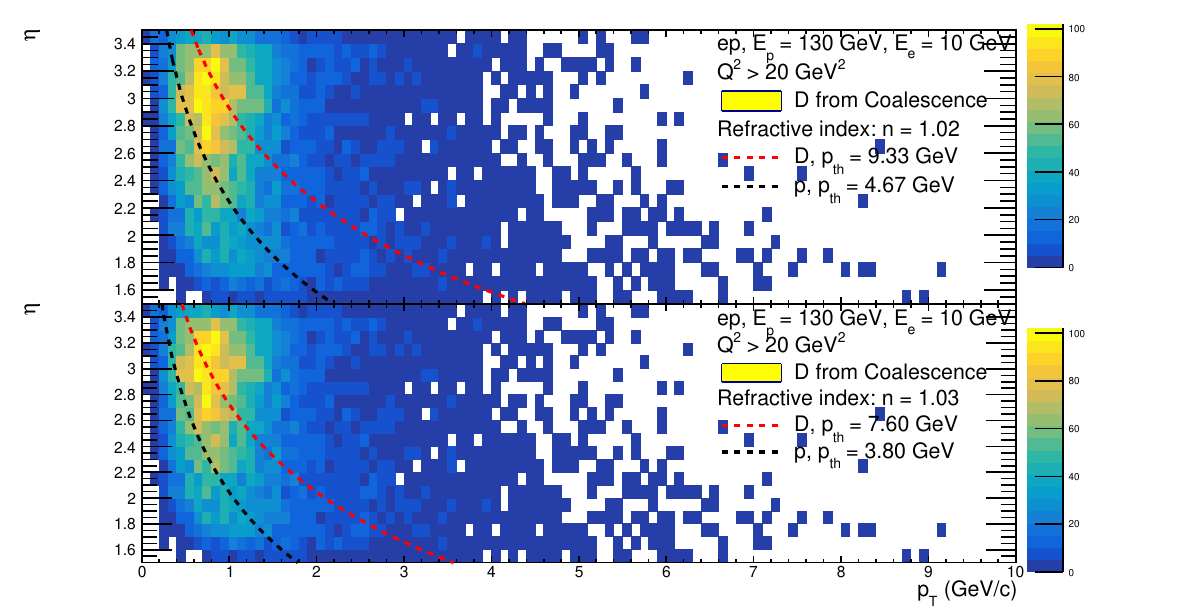}
        \caption{Transverse-momentum versus pseudorapidity distribution of deuterons produced via coalescence in ep collisions within the ePIC dRICH acceptance. The top and bottom panels correspond to aerogel refractive indices $n=1.02$ and $n=1.03$, respectively. Dashed lines indicate the proton and deuteron Cherenkov thresholds. Above the deuteron threshold, identification is based on the Cherenkov angle; between the two thresholds, a hadron-veto technique can be used.}
        \label{fig:drichpid}
    \end{minipage}
    \hfill
    \begin{minipage}[t]{0.39\textwidth}
        \centering
        \includegraphics[width=\textwidth]{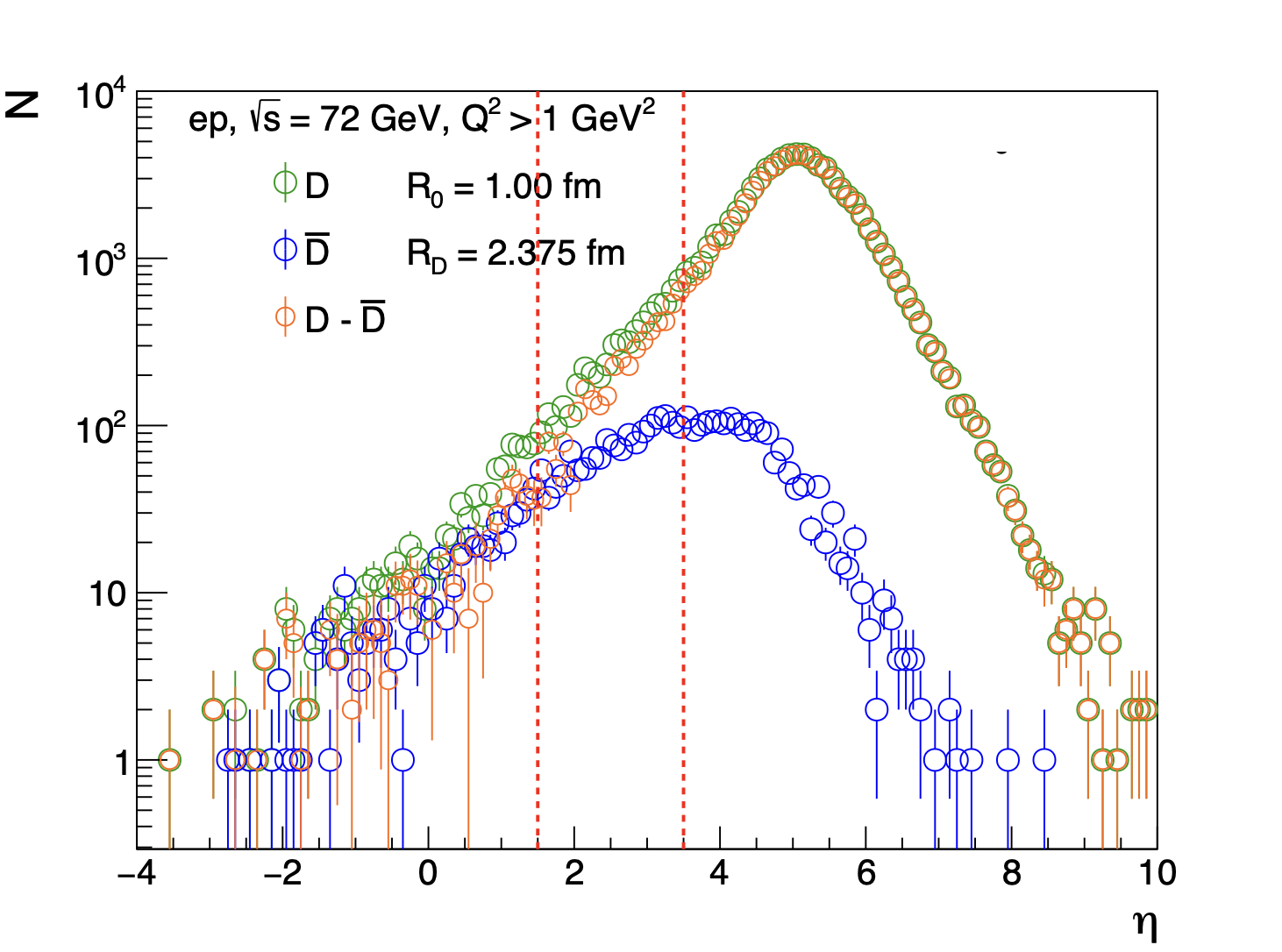}
        \caption{Predicted deuteron, antideuteron, and difference yields as a function of pseudorapidity in ep collisions at $\sqrt{s}=72$~GeV for $Q^{2}>1$~GeV$^{2}$. Dashed vertical lines indicate the ePIC dRICH acceptance, $1.5 < \eta < 3.5$. Simulation: $10^{9}$ ep collisions at $E_{p}=130$~GeV and $E_{e}=10$~GeV generated with \textsc{Pythia8} Monash and processed with the coalescence afterburner, using $R_{0}=1.00$~fm and $R_{D}=2.375$~fm.}
        \label{fig:eicprediction}
    \end{minipage}
\end{figure}

Figure~\ref{fig:eicprediction} shows the predicted deuteron and antideuteron yields as a function of pseudorapidity. A clear asymmetry is observed, with a larger deuteron yield in the proton-going direction. This behaviour originates from the valence-quark structure of the proton and makes the measurement potentially sensitive to proton parton distribution functions.

The region $1.5 < \eta < 3.5$, highlighted in Fig.~\ref{fig:eicprediction}, corresponds to the ePIC dRICH acceptance. Within this region, the model predicts about $6\times10^{5}$ deuterons and $1.2\times10^{5}$ antideuterons in the first year of data taking. The forward region is therefore particularly interesting for the expected yields and the sensitivity of the nuclei--antinuclei asymmetry to proton structure.

Deuteron and antideuteron identification with the ePIC dRICH is illustrated in Fig.~\ref{fig:drichpid}. The detector covers $1.5<\eta<3.5$ and employs an aerogel radiator with refractive index $n=1.02$--$1.03$. For $n=1.02$, the proton and deuteron Cherenkov momentum thresholds are  $4.67$~GeV/$c$ and $9.33$~GeV/$c$, while for $n=1.03$ they decrease to about $3.80$~GeV/$c$ and $7.60$~GeV/$c$.

Two particle-identification regimes can be exploited. Above the deuteron Cherenkov threshold, deuterons and antideuterons can be identified through the Cherenkov angle. Between the proton and deuteron thresholds, identification can rely on a veto technique: Cherenkov emission indicates compatibility with a proton (or a lower mass particle), while its absence for a reconstructed charged track is consistent with a deuteron.

\section{Conclusions}

A coalescence model coupled to \textsc{Pythia8} has been validated against ALICE pp measurements and ZEUS ep data, showing a consistent description of deuteron and antideuteron production across collision systems. At EIC energies, the model predicts large deuteron and antideuteron samples within the ePIC dRICH acceptance and a pronounced matter--antimatter asymmetry in the proton-going region. Light-nuclei production can therefore provide a new observable for testing coalescence and probing proton structure at the EIC.

\end{document}